\documentclass[conference,USletter,10pt]{IEEEtran}
\usepackage{graphicx} 
\usepackage{amsmath}
\usepackage{amsfonts}
\usepackage{amssymb}
\usepackage[ruled,vlined,linesnumbered]{algorithm2e}
\usepackage{xcolor}

\SetCommentSty{mycommfont}

\ifCLASSINFOpdf

\else

\fi

\begin{document}

\title{Backhaul Traffic Balancing and Dynamic Content-Centric Clustering for the Downlink of Fog Radio Access Network}
\author{\IEEEauthorblockN{Di Chen, Stephan Schedler and Volker Kuehn}
\IEEEauthorblockA{Institute of Communications Engineering\\
University of Rostock \\
Email: \{di.chen, stephan.schedler, volker.kuehn\}@uni-rostock.de}
}

\maketitle

\begin{abstract}
Recently, an evolution of the Cloud Radio Access Network (C-RAN) has been proposed, named as Fog Radio Access Network (F-RAN). Compared to C-RAN, the Radio Units (RUs) in F-CAN are equipped with local caches, which can store some frequently requested files. In the downlink, users requesting the same file form a multicast group, and are cooperatively served by a cluster of RUs. The requested file is either available locally in the cache of this cluster or fetched from the Central Processor (CP) via backhauls. Thus caching some frequently requested files can greatly reduce the burden on backhaul links. Whether a specific RU should be involved in a cluster to serve a multicast group depends on its backhaul capacity, requested files, cached files and the channel. Therefore it is subject to optimization. In this paper we investigate the joint design of multicast beamforming, dynamic clustering and backhaul traffic balancing. Beamforming and clustering are jointly optimized in order to minimize the power consumed, while QoS of each user is to be met and the traffic on each backhaul link is balanced according to its capacity.
\end{abstract}

\IEEEpeerreviewmaketitle

\section{Introduction}
The evolution toward 5G is featured by the explosive growth of traffic in the network, due to the exponentially increased number of user terminals and QoS demands. Moreover, besides Spectral Efficiency (SE), Energy Efficiency (EE) also becomes an important metric for the design of 5G system in order to decrease the global Carbon Dioxide $(\mathrm{CO}_2)$ emissions and operational costs. Compared to 4G system, it is widely recognized that 5G should achieve the growth by a factor of 1000 in terms of system capacity and a factor of 10 in EE. Several approaches have been considered to achieve this objective, e.g., increase of frequency spectrum usage, increase of both per-link  and area spectral efficiency. Millimeter Wave (mm Wave) communication at 28 GHz and 60 GHz is currently studied worldwide to overcome the shortage of frequency resource. In order to increase the per-link spectral efficiency, C-RAN \cite{CRAN} has been shown to be a promising architecture for its much more efficient interference management due to the centralized processing. A straightforward way to increase the area spectral efficiency is to decrease the distance of transmitters and receivers. This can be done by densely deploying low-cost Radio Units (RU). A crucial problem of this approach is that densely deployed of low-cost RUs usually connect to the CP via backhauls with very limited capacities. Thus the backhauls between RUs and the CP in the cloud become the main bottleneck of the overall performance of the network. Introducing local cache to RU has shown to be a possible solution to this problem.  It has been demonstrated in \cite{CacheAir} that with edge caching, overhead can be greatly reduced and higher spectral efficiency and lower latency can be obtained. An evolution of C-RAN named as Fog Radio Access Network (F-RAN) has been recently studied in \cite{F-CAN} by adopting this idea. When some requested files are cached locally at RUs, there is no need to fetch them remotely from the CP via backhauls, thus the burden on backhaul is relieved and its capacity is not the performance bottleneck for users requesting cached files. By incorporating caching units at RUs, F-RAN with densely deployed low-cost RUs can ease some difficulties of C-RAN and increase both per-link and area spectral efficiency.

Recent studies have shown that popular multimedia streaming produces a significant portion of traffic, e.g., newly-released movies or live sport matches. Unequal popularity makes caching popular files more meaningful and sensible, which can greatly relieve the burden on backhauls and reduce the delay for a large number of users. In \cite{FundLimCach}, the information theoretical fundamental limits of caching in a broadcast channel are characterized and two caching schemes are proposed, i.e., uncoded caching and coded caching. In uncoded caching, complete files are cached, while in coded caching, different fractions (e.g. parity bits) of the files can be
stored at different caches using MDS codes (e.g. Fountain code). In \cite{CCMBF} and \cite{MeixiaCach}, trade-off between the total power consumed and the total backhaul capacity needed in the downlink of F-RAN is studied. The author assumes that uncoded caching is used, and each RU caches the same most popular files until the caching memory is full. For users requesting cached files, RUs can cooperatively serve these users without consuming the backhaul resource. For users requesting uncached files, remote delivery of these files from the CP to RUs will consume backhaul resource. When more RUs are involved in the cluster for transmitting uncached files, power consumption will be decreased due to the increase of spatial diversity by cooperative communication, while the burden on backhaul is increased due to the delivery of files to more RUs. When less RUs are involved, the decrease of backhaul costs will lead to higher power consumption due to less cooperation. This trade-off is characterized in \cite{MeixiaCach} for uncoded caching.  A similar trade-off for coded caching is characterized in \cite{CodedCache}. However, all these works emphasize the trade-off between total power and total backhaul capacity, which may result in severe imbalanced data traffic on backhauls. In this paper, we address this problem by considering an individual backhaul capacity associated with each RU and try to minimize the power consumption for higher EE. We propose an efficient algorithm, which
\begin{itemize}
	\item  dynamically nulls out the RUs with little contribution in a cluster serving the corresponding multicast group, in order to reduce the backhaul cost and satisfy individual capacity constraints;
	\item retains the RU with large contribution to this multicast group in order to guarantee the QoS ;
	\item  designs optimal beamforming vectors so as to minimize the power consumption.
\end{itemize}

This paper is structured as follows:  In Sec. \uppercase\expandafter{\romannumeral2} we introduce the channel model considered and state the problem mathematically. The optimization algorithm is presented and explained in Sec. \uppercase\expandafter{\romannumeral3}. Simulation results and conclusions are provided in Sec. \uppercase\expandafter{\romannumeral4} and Sec. \uppercase\expandafter{\romannumeral5} respectively.
\section{System Model and Problem Statement}
\begin{figure}
	\centering\includegraphics[width=0.48\textwidth]{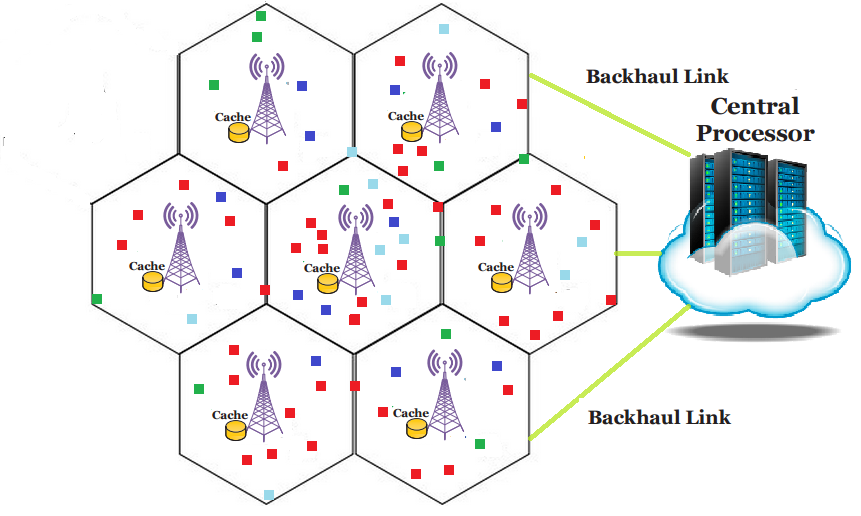}
	\caption{Channel Model: Downlink of F-RAN \cite{MeixiaCach}.}\label{fig:FCAN} 
\end{figure}
\subsection{System Model} 
We consider the downlink transmission of a hexagonal multi-cell F-RAN as illustrated in Fig \ref{fig:FCAN}.  $N$ Radio Units (RUs) are located in the network and cooperatively serve all users. Let $\mathcal{N}=\{1,2,...,N\}$ denote the set of RUs. Each RU is located at the center of a hexagonal-type cell and equipped with $L$ antennas and a cache. It connects to the Central Processor (CP) in the cloud via individual backhaul links with finite capacity $C_{\mathrm{BH},i}$.  $d_{\mathrm{RU}}$ denotes he distance between adjacent RUs. The colored square dots denote single-antenna users, which are uniformly and independently distributed within the network. In each scheduling interval, $K$ users will be scheduled, and send their content request according to certain demand probability. Users with same color request the same file $f_m$ and form a multicast group $\mathcal{G}_m$. Let $M$ denote the number of different multicast groups and assume that each user can request at most one content at its scheduled time. Hence, $\mathcal{G}_i\cap\mathcal{G}_j=\emptyset,\ \forall i\ne j$, and $\sum_{m=1}^{M}|\mathcal{G}_m|\leq K$ hold. The $m$-th multicast group $\mathcal{G}_m$ is served cooperatively by a cluster of RUs, denoted by $\mathcal{C}_m$ and $\mathcal{C}_m\subseteq\mathcal{N}$. Unlike the multicast group $\mathcal{G}_m$, which is fixed based on users' requests, its cluster of serving RUs $\mathcal{C}_m$ is subject to be dynamically optimized by the CP, and they can overlap with each other, i.e., $\mathcal{C}_i\cap\mathcal{C}_j$ is not necessary an empty set. Moreover, we assume that CSI is available to the CP and the channel is block fading, which remains constant within a time scheduling but can change from one frame to another.

Let $s_m$ denote the transmitted symbol from file $f_m$ to the $m$-th multicast group $\mathcal{G}_m$ with normalized power $\mathbb{E}\{|s_m|^2\}=1,\ \forall m\in[1,...,M]$. This symbol will be transmitted cooperatively by all RUs in cluster $\mathcal{C}_m$. Suppose that user $k$ is in the $m$-th multicast group $\mathcal{G}_m$, i.e., $k\in\mathcal{G}_m$. The channel vector from the $n$-th RU to the $k$-th user is denoted by $\mathbf{h}_{k,n}\in\mathbb{C}^{L\times 1}$. Thus, the aggregated channel vector from all RUs to user $k$ is denoted as $\mathbf{h}_{k}=[\mathbf{h}_{k,1}^H,\mathbf{h}_{k,2}^H,...,\mathbf{h}_{k,n}^H]^H\in\mathbb{C}^{NL\times 1}$. For the multicast scenario considered here, the beamforming vector construction is content-centric \cite{CCMBF}, such that it is based on the transmitted content, or equally, its served multicast group. Assume that at RU $n$, the beamforming vector for $m$-th multicast group $\mathcal{G}_m$ is denoted by $\mathbf{v}_{m,n}\in\mathbb{C}^{L\times 1}$. Thus, $\mathbf{v}_{m,n}=\mathbf{0}_{L\times 1}$ means that RU $n$ is not involved in cluster $\mathcal{C}_m$ to serve multicast group $\mathcal{G}_m$. Then, the aggregated beamforming vector from all RUs for multicast group $\mathcal{G}_m$ is $\mathbf{v}_{m}=[\mathbf{v}_{m,1}^H,\mathbf{v}_{m,2}^H,...,\mathbf{v}_{m,n}^H]^H\in\mathbb{C}^{NL\times 1}$. Hence, the SINR at user $k$ can be expressed as
\begin{equation}\label{SINRorig}
\mathrm{SINR}_k=\frac{|\mathbf{h}_{k}^H\mathbf{v}_{m}|^2}{\sigma_k^2+\sum_{i\ne m}^{M}|\mathbf{h}_{k}^H\mathbf{v}_{i}|^2},\qquad k\in\mathcal{G}_m,
\end{equation}
where $\sigma_k^2$ denotes variance of the i.i.d additive complex Gaussian noise with zero mean at user $k$.
\subsection{File and Cache Model}
We assume that all $F$ files are available at the CP in the cloud, have the same normalized length 1 but different popularities. Without loss of generality, the files are indexed in the order from the most to the least popular ones, such that the most popular file has index $f=1$ and the least popular file has index $f=F$. The popularity of the file is modeled by Zipf distribution \cite{FemtoCache}, i.e., the probability of a file $f$ is requested is
\begin{equation}\label{Zipf}
P(f)=\frac{f^{-\alpha}}{\sum_{j=1}^{F}j^{-\alpha}},\qquad f=\{1,2,...,F\}.
\end{equation}
$\alpha$ is related to the skewness of the distribution, larger $\alpha$ makes the probability of
requesting a small group of files larger.

We adopt the same cache strategy and model as in \cite{CCMBF} and \cite{MeixiaCach}, where each RU is equipped with a cache of memory $S$, and uncoded caching is utilized in order to increase the probability of cooperation, thus lower the power consumption. Files with index smaller than or equal to $S$ are cached at all RUs. Note that the caching strategy and content are fixed and the caching placement problem is not addressed here, which is beyond of the scope of this paper. In each scheduled interval, each scheduled user sends a file request, which fulfills the Zipf distribution \eqref{Zipf}. The cached files are transmitted directly from RUs without consuming the backhaul resource. Contrarily, the uncached files must be fetched remotely from the CP to all RUs in the cluster serving this multicast group. Obviously, compared to coded caching, in uncoded caching strategy, the RUs in a specific cluster cooperatively transmit the same content, the increase of spatial diversity leads to more load on the backhaul. Thus traffic handling is a significant issue, especially for uncoded caching, which is not addressed in \cite{CCMBF}, \cite{MeixiaCach} and \cite{CodedCache}, and it will be addressed below.

\subsection{Problem Statement}
In this paper, we aim to minimize the power consumption, while the QoS of each user has to be met and the backhaul traffic must be balanced according to individual capacities. The problem is formulated as follows:
 \begin{align}
&\mathcal{P}_{\mathrm{Original}}:\qquad\mathop {\min}\limits_{\mathbf{v}_{m,n}}\qquad\sum_{m=1}^{M}||\mathbf{v}_{m}||_2^2\label{proborig}\\ 
&\mathrm{s.t.}\qquad \mathrm{SINR}_k\geq\Gamma_m,\quad \forall k\in\mathcal{G}_m,\quad\forall \mathcal{G}_m;\label{SINRconstr}\\
&\sum_{m=1}^{M}(1-c_{f_m,n})\big|||\mathbf{v}_{m,n}||_2^2\big|_0\log_2\left(1+\Gamma_m \right)\leq C_{\mathrm{BH},n}, \forall n\in\mathcal{N}.\label{BHconstr}
\end{align}
Eq. \eqref{proborig} is the total power consumed, which is the sum of the power consumed for each multicast group among all RUs. If RU $n$ is not involved in cluster $\mathcal{C}_m$ to serve users in multicast group $\mathcal{G}_m$, the corresponding beamforming vector $\mathbf{v}_{m,n}$ will be the zero vector. Constraint \eqref{SINRconstr} guarantees the QoS of each user in each multicast group, where $\Gamma_m$ denotes the target SINR of the content requested by $\mathcal{G}_m$ and $ \mathrm{SINR}_k$ is defined in \eqref{SINRorig}. Constraint \eqref{BHconstr} guarantees the traffic on each backhaul does not exceed its capacity, where $c_{f_m,n}\in\{0,1\}$. $c_{f_m,n}=1$ denotes the requested file $f_m$ of multicast group $\mathcal{G}_m$ is cached at $n$-th RU, i.e., $f_m\leq S$, otherwise it is zero. $c_{f_m,n}$ is fixed and known to the CP once current scheduled users have submitted their requests. We use $\ell_0$-norm to denote whether beamforming vector $\mathbf{v}_{m,n}$ is zero vector or not, i.e., when $n$-th RU involves in cluster $\mathcal{C}_m$ serving $\mathcal{G}_m$, $\big|||\mathbf{v}_{m,n}||_2^2\big|_0=1$, otherwise it is zero. We see that the backhaul of RU $n$ is used for $\mathcal{G}_m$ only if the requested file is not cached $(1-c_{f_m,n}=1)$ and it contributes to cluster $\mathcal{C}_m$ $(\big|||\mathbf{v}_{m,n}||_2^2\big|_0=1)$. In this case the resource consumed on the backhaul is $\log_2\left(1+\Gamma_m \right)$ at minimum, when Gaussian codebooks are used. By summing up all multicast groups, we obtain the total backhaul resource consumption at RU $n$ in \eqref{BHconstr}, which should be smaller than its capacity. 

The descriptions above shows that clustering and beamforming vectors are closely related to requested files, cached files, individual backhaul link capacities and the channel between all RUs and scheduled users. For different scheduling intervals, the above parameters (except for backhaul capacities) change independently and dynamically, thus an efficient optimization scheme is necessary. Although we do not explicitly optimize the clustering scheme, it is implicitly optimized and determined by the resulting $\big|||\mathbf{v}_{m,n}||_2^2\big|_0$ of the problem.
\section{Optimization Problem}
In this section we illustrate the procedures to solve problem \eqref{proborig}. Due to the non-convex constraints \eqref{SINRconstr} and \eqref{BHconstr}, the problem is in general non-convex. Hence, in the following 2 subsections, we adopt two techniques to convexify them.
\subsection{SDR: Convexification of SINR constraints \eqref{SINRconstr}}
From CP's perspective, the downlink of F-RAN is actually a \textit{virtual} multi-antenna multicast system, which is similar to the problem solve in \cite{QoSMMBF}. In that work, in order to convexify the similar SINR constraints, a Semi-Definite Relaxation (SDR) technique is proposed. We also adopt this idea in our problem.

Let $\mathbf{V}_m=\mathbf{v}_m\mathbf{v}_m^H$ and $\mathbf{H}_k=\mathbf{h}_k\mathbf{h}_k^H$, $\forall m,k$, where both $\mathbf{V}_m, \mathbf{H}_k\in\mathbb{C}^{NL\times NL}$ are positive semidefinite matrices. We define the selection matrix at RU $n$ as $\mathbf{J}_n=\mathrm{Diag}\left(\left[\mathbf{0}_{(n-1)L\times 1}^H,\mathbf{1}_{L\times 1}^H,\mathbf{0}_{(N-n)L\times 1}^H\right]\right)$, and thus $||\mathbf{v}_{m}||_2^2=\mathrm{tr}(\mathbf{V}_m), ||\mathbf{v}_{m,n}||_2^2=\mathrm{tr}(\mathbf{V}_m\mathbf{J}_n),$ and $ |\mathbf{h}_{k}^H\mathbf{v}_{m}|^2=\mathrm{tr}(\mathbf{V}_m\mathbf{H}_k)$. Then \eqref{proborig} and \eqref{SINRconstr} can be equivalently expressed as: 
 \begin{align}
&\mathop {\min}\limits_{\mathbf{V}_m}\qquad\sum_{m=1}^{M}\mathrm{tr}(\mathbf{V}_m)\label{probsdr}\\ 
&\mathrm{s.t.}\ \Gamma_m\left(\sigma_k^2+\sum_{i\ne m}^{M}\mathrm{tr}(\mathbf{V}_i\mathbf{H}_k)\right)-\mathrm{tr}(\mathbf{V}_m\mathbf{H}_k)\leq 0, \forall k;\label{SINRconstrsdr}\\
&\mathbf{V}_m\succeq \mathbf{0}.\ \forall m=\{1,2,...,M\};\label{psd}\\
&\mathrm{rank}\left(\mathbf{V}_m\right)=1,\ \forall m=\{1,2,...,M\}.\label{rank1}
\end{align}
The relaxation step of SDR technique is to drop the last rank-one constraints, which are non-convex, and solve the relaxed convex problem \eqref{probsdr}-\eqref{psd}. It is a standard Semi-Definite Programming (SDP) problem. If the obtained optimal $\mathbf{V}_m$ has rank 1, the EigenValue Decomposition (EVD) can be used to obtain the corresponding optimal beamforming vector $\mathbf{v}_m$. Otherwise randomization and scaling method is used to generate a
suboptimal solution. Details can be found in \cite{QoSMMBF}. 
\subsection{Re-weighted $\ell_1$-norm: Convexification of backhaul constraints \eqref{BHconstr}}
Constraint \eqref{BHconstr} is non-convex due to the discrete $\ell_0$-norm $\big|||\mathbf{v}_{m,n}||_2^2\big|_0$. We adopt the technique proposed in a compressive sensing literature \cite{l1norm}, which is also adopted in \cite{EEinDownCRAN}, to convexify it. $\ell_0$-norm is iteratively approximated by a re-weighted $\ell_1$-norm, which is linear, continuous and convex, i.e.,
\begin{equation}\label{l1approx}
\begin{aligned}
&\big|||\mathbf{v}_{m,n}^{(t+1)}||_2^2\big|_0=\big|\mathrm{tr}(\mathbf{V}_m^{(t+1)}\mathbf{J}_n)\big|_0\approx w_{m,n}^{(t+1)}\mathrm{tr}(\mathbf{V}_m^{(t+1)}\mathbf{J}_n)\\
&\mathrm{with}\  w_{m,n}^{(t+1)}=\frac{1}{\tau+\mathrm{tr}(\mathbf{V}_m^{(t)}\mathbf{J}_n)},\ \tau>0.
\end{aligned}
\end{equation}
For clarity, at first we drop the superscript $(t)$ and $(t+1)$ in \eqref{l1approx} to explain the approximation. We see that $\big|\mathrm{tr}(\mathbf{V}_m\mathbf{J}_n)\big|_0$ is approximated as $w_{m,n}\mathrm{tr}(\mathbf{V}_m\mathbf{J}_n)=\frac{\mathrm{tr}(\mathbf{V}_m\mathbf{J}_n)}{\tau+\mathrm{tr}(\mathbf{V}_m\mathbf{J}_n)}$. When $\mathrm{tr}(\mathbf{V}_m\mathbf{J}_n)\gg\tau$, this linear weighted approximation of $\ell_0$-norm is close to 1. Contrarily, the approximation quickly approaches 0 for $\mathrm{tr}(\mathbf{V}_m\mathbf{J}_n)\ll\tau$. Thus, $\tau$ can be regarded as a parameter that defines the threshold which determines whether $\mathrm{tr}(\mathbf{V}_m\mathbf{J}_n)$ is \textit{on} (1) or \textit{off} (0). Hence, this continuous and linear approximation captures the behavior of discrete non-convex $\ell_0$-norm by carefully selecting the value of $\tau$. Now we add the superscripts in \eqref{l1approx}, which describes the iterative re-weighted procedure. Once an updated $\mathbf{V}_m^{(t)}$ is obtained from the $t$-th iteration, the weighted coefficient $w_{m,n}^{(t+1)}$ of approximation used for the $(t+1)$-th iteration should also be updated for a precise approximation. In the iterative procedure, at RU $n$, if the transmit power $\mathrm{tr}(\mathbf{V}_m\mathbf{J}_n)$ for multicast group $\mathcal{G}_m$ decreases, it would have higher weight $w_{m,n}$ in the next iteration, then its value would be forced to further reduce and encouraged to drop out of being involved in this cluster eventually, in order to relieve the burden on backhaul.

\subsection{Reformulation and solution of original problem \eqref{proborig}-\eqref{BHconstr}}
By combining the relaxation and approximation techniques of Sec.\uppercase\expandafter{\romannumeral3}.A and Sec.\uppercase\expandafter{\romannumeral3}.B, we can reformulate our original problem $\mathcal{P}_{\mathrm{Original}}$ as follows:
{\small \begin{align}
&\mathcal{P}_{\mathrm{Ref}}:\qquad\mathop {\min}\limits_{\mathbf{V}_m}\qquad\sum_{m=1}^{M}\mathrm{tr}(\mathbf{V}_m)\label{probsdrref}\\ 
&\mathrm{s.t.}\ \Gamma_m\left(\sigma_k^2+\sum_{i\ne m}^{M}\mathrm{tr}(\mathbf{V}_i\mathbf{H}_k)\right)-\mathrm{tr}(\mathbf{V}_m\mathbf{H}_k)\leq 0, \forall k;\label{SINRconstrsdrref}\\
&\sum_{m=1}^{M}R_{m,n}w_{m,n}\mathrm{tr}(\mathbf{V}_m\mathbf{J}_n)-C_{\mathrm{BH},n}\leq 0,\ \forall n\in\mathcal{N};\label{BHconstrref}\\
&\mathbf{V}_m\succeq \mathbf{0},\ \forall m=\{1,2,...,M\}\label{psdref}.
\end{align}}
$R_{m,n}=(1-c_{f_m,n})\log_2(1+\Gamma_m)$ is constant in each scheduled interval and known to the CP. The reformulated problem consists of only a linear objective function, $K+N$ linear
inequality constraints and $M$ positive-semidefinite constraints, which is a standard SDP problem and can be efficiently solved by many solvers, such as SDPT3 and SeDuMi. In order to have a better and precise approximation of the $\ell_0$-norm and null out small powers eventually, we solve $\mathcal{P}_{\mathrm{ref}}$ iteratively, with updated coefficient $w_{m,n}$ based on the solution of the previous iteration. For the initial value of the weighted coefficient- $w_{m,n}^{(1)}=\frac{1}{\mathrm{tr}(\mathbf{V}_m^{(0)}\mathbf{J}_n)+\tau}$, we solve a \textbf{initial problem} $\mathcal{P}_{\mathrm{Init}}$, which \textbf{excludes} the constraint \eqref{BHconstrref} in $\mathcal{P}_{\mathrm{Ref}}$, to obtain the initial value of $\mathrm{tr}(\mathbf{V}_m^{(0)}\mathbf{J}_n)$. When optimal $\mathbf{V}_m$ is obtained after the last iteration, EVD or randomization and scaling method is utilized to obtain its corresponding beamforming vector $\mathbf{v}_m$, as shown in \cite{QoSMMBF}. The overall algorithm is listed below: 
\begin{algorithm}
	{\textbf{Initialization}: Solve standard SDP problem $\mathcal{P}_{\mathrm{Init}}$ to obtain $\mathbf{V}_m^{(0)}$. Compute $w_{m,n}^{(1)}$ based on \eqref{l1approx}, $\forall m,n$. Construct problem $\mathcal{P}_{\mathrm{Ref}}^{(1)}$, set $t\leftarrow 1$.\\
	 \Repeat{Convergence or reaching max iteration number}{Solve standard SDP problem $\mathcal{P}_{\mathrm{Ref}}^{(t)}$ to obtain $\mathbf{V}_m^{(t)}$\\ Update $w_{m,n}^{(t+1)} $ based on \eqref{l1approx}, then update problem $\mathcal{P}_{\mathrm{Ref}}^{(t+1)}$, set $t\leftarrow t+1.$} 
	 \If{$\mathrm{rank}(\mathbf{V}_m^{\mathrm{last}})=1$}{Perform EVD to obtain optimal $\mathbf{v}_m$.}
	 \Else{Use Gaussian randomization and scaling \cite{QoSMMBF} to obtain the approximate solution $\mathbf{v}_m^*$.}	
		\caption{Iterative Optimization Steps \label{Alg}}}
\end{algorithm}\label{Alg}
\section{Simulation Results}
In this section, simulation results based on the proposed algorithm are provided. The model in Fig. \ref{fig:FCAN} is considered, with simulation parameters listed in Table on the right. Most results are based on these parameters unless otherwise stated.

\begin{table}\label{table}
	\caption{Simulation parameters}
\begin{center}
\begin{tabular}{| c | c |}
		\hline
		Number of RU (Hexagonal Cell): $N$ & 7 \\ \hline
		Equipped antennas of each RU: $L$ & 2 \\ \hline
		Distance between adjacent RUs: $d_{\mathrm{RU}}$ & 0.5 km \\ \hline
	    Transmit Antenna Gain  & 10 dBi \\ \hline
	    Total number of users: $K_{\mathrm{tot}}$ & 200 \\ \hline
	    Number of scheduled users per interval: $K$ & 12 \\ \hline
	    Background noise  & -172 dBm/Hz \\ \hline
	    3GPP LTE-A path loss model  & $148.1+37.6\log_{10}(d)$ \\ \hline
	    Log-normal shadowing  & 8 dB \\ \hline
	    Rayleigh small scale fading  & 0 dB \\ \hline
	    Bandwidth: $B$  & 10 MHz \\ \hline
	    Target SINR at each user: $\Gamma$  & 10 dB \\ \hline
	    Total number of files: $F$ & 100 \\ \hline
	    Skew parameter of Zipf distribution: $\alpha$ & 1.5 \\ \hline
	    Cache Memory: $S$ & 3 \\ \hline
	    Individual backhaul capacity: $C_{\mathrm{BH}}$ & 70Mbps \\ \hline
	    Threshold parameter in \eqref{l1approx}: $\tau$ & -50dBm\\ \hline
\end{tabular}
\end{center}
\end{table}
In the first realization of simulation, after 12 scheduled users submit their requests, the CP knows that totally 7 files are requested, thus 7 multicast groups are formed. Only 2 of them have been cached at all RUs this time, whose index is 1 and 3 respectively. Although cache memory $S$ is 3, the cached file with index 2 is not requested in this realization. Without loss of generality, we name the two requested files that are cached as $\mathrm{f}(1)$ and $\mathrm{f}(2)$ for multicast group 1 and 2, respectively.
Then we run the proposed algorithm \ref{Alg} and the algorithm in \cite{MeixiaCach}, and record the power $P_{m,n}=||\mathbf{v}_{m,n}||_2^2$ allocated to file $f(m)$ requested by multicast group $m$ at each RU over each iteration, so as to see the resultant clustering pattern. 

\begin{figure}
	\centering\includegraphics[width=0.48\textwidth]{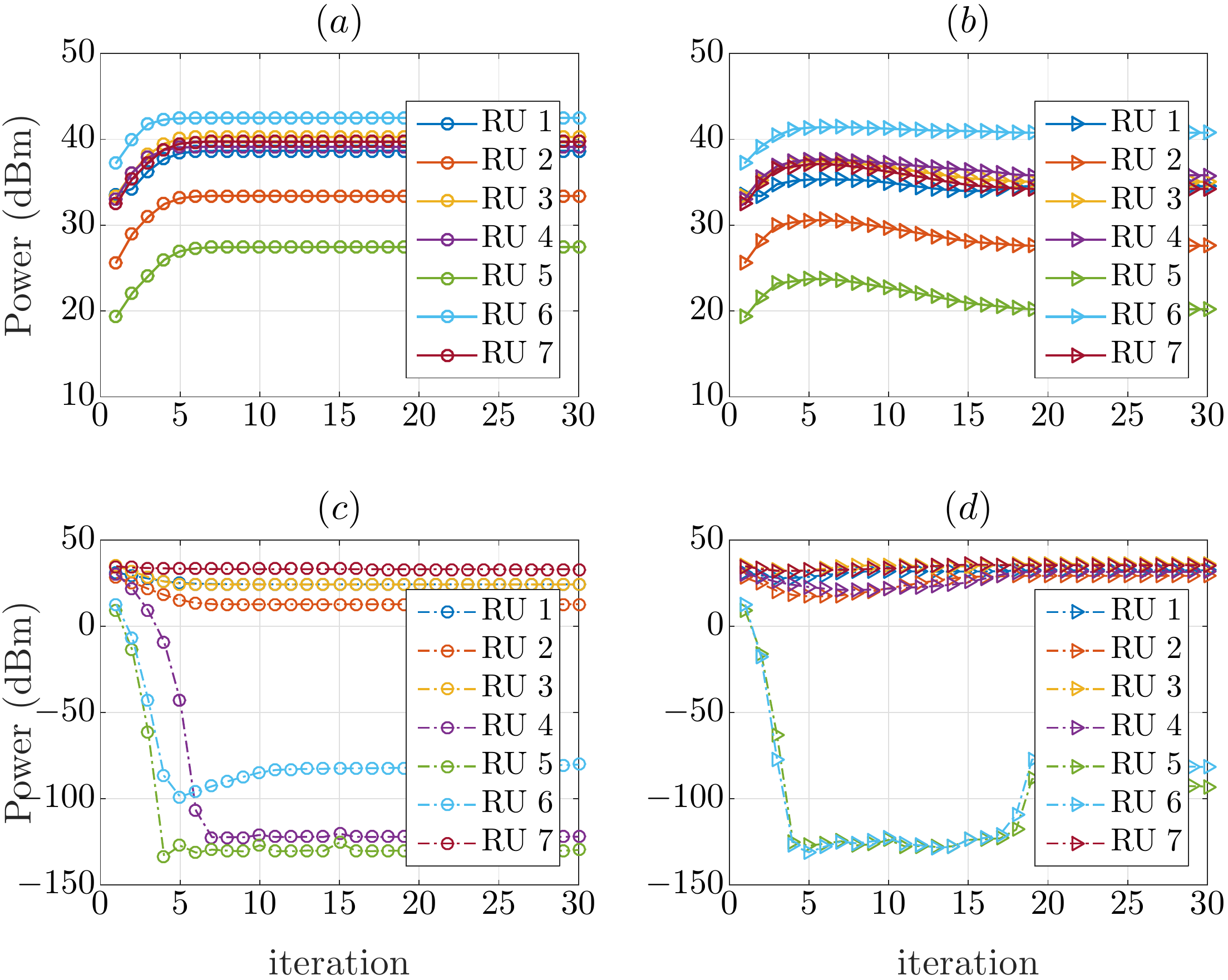}
	\caption{Clustering for cached File $f(2)$: $(a),(b)$ and uncached File $f(6)$: $(c),(d)$ at 7 RUs. Proposed Alg.: $(a),(c)$; Alg. in \cite{MeixiaCach}: $(b),(d)$.}\label{file}
\end{figure}

In Fig. \ref{file}, the clustering pattern is illustrated from requested file's perspective. $(a),(b)$ show the allocated power for cached file $\mathrm{f}(2)$ at all 7 RUs, plotted as solid lines. $(c),(d)$ show it for uncached file $\mathrm{f}(6)$, plotted as dotted-dashed lines. The figures $(a),(c)$ on left side with circle markers are obtained by running the proposed algorithm. The figures $(b),(d)$ on right side with triangle markers are obtained with algorithm in \cite{MeixiaCach}. Note that the threshold is set as $-50$ dBm, for cached file $f(2)$, all 7 RUs in both algorithm will participating in transmitting this file. Since cached files do not consume backhaul resource, involving all RUs in this cluster can always increase spatial diversity and thus decrease power consumption. Hence, for cached files, the clustering results are the same for both algorithm. This result is in consistence with the theory proposed in \cite{MeixiaCach}. Moreover, our proposed algorithm produces stable results just after about 5 iterations. However, the results for uncached file $f(6)$ are different. Since backhaul resource is consumed, involvement of all RU in cluster for these files is not possible. In $(c)$, our proposed algorithm nulls out 3 RUs after 7 iterations in cluster for transmitting $f(6)$, in order to meet each individual backhaul capacity constraint. While only 2 RUs for this cluster are nulled out by algorithm in \cite{MeixiaCach}, which might cause traffic problems, as we show next.

\begin{figure}
	\centering\includegraphics[width=0.48\textwidth]{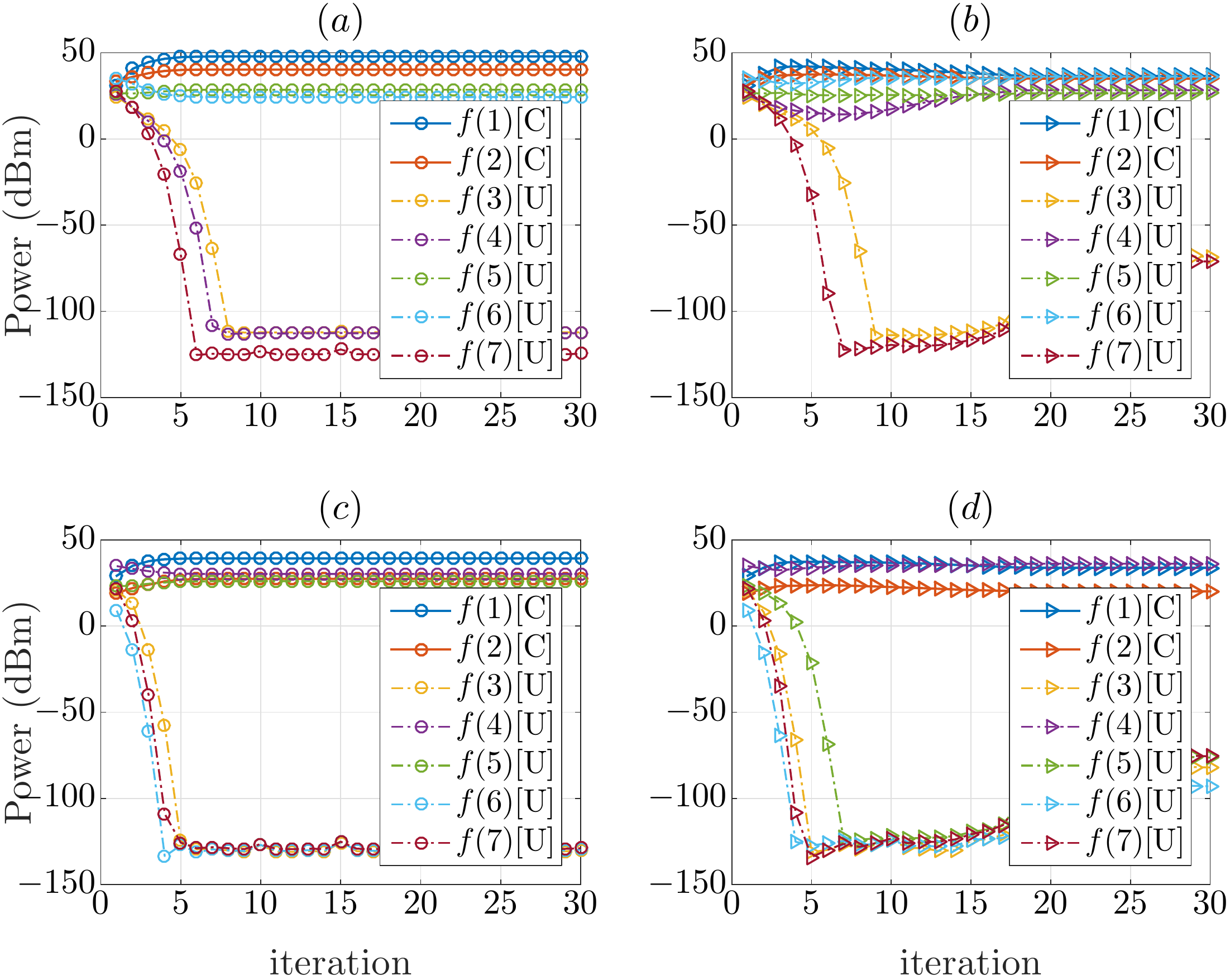}
	\caption{Clustering for all cached [C] and uncached [U] files at RU 3: $(a),(b)$ and RU 5: $(c),(d)$. Proposed Alg.: $(a),(c)$; Alg. in \cite{MeixiaCach}: $(b),(d)$.}\label{ru}
\end{figure}

In Fig. \ref{ru}, we illustrate the clustering from the RU's perspective. Each RU might participate in several clusters. The backhaul capacity of each RU is $C_{\mathrm{BH}}=70$ Mbps. Thus besides supporting 2 cached files without consuming backhaul resources, each RU can support at most 2 uncached data streams since $B\log_2(1+\Gamma)\times 2\approx 70$ Mbps. $(a),(c)$ show that with the proposed algorithm, the clustering pattern is constructed such that exactly 2 data streams of uncached files are transmitted at RU 3 and 5. With the algorithm in \cite{MeixiaCach}, RU 3 has to support 3 data streams of uncached files and RU 5 supports only 1. Although this is optimal if all RUs share a common backhaul resource, which is not always the case in practice. Individual backhauls (e.g. optical fibers) are usually predetermined. Hence, the clustering pattern obtained by \cite{MeixiaCach} may cause traffic congestion and resource waste in practice.

\begin{figure}
	\centering\includegraphics[width=0.48\textwidth]{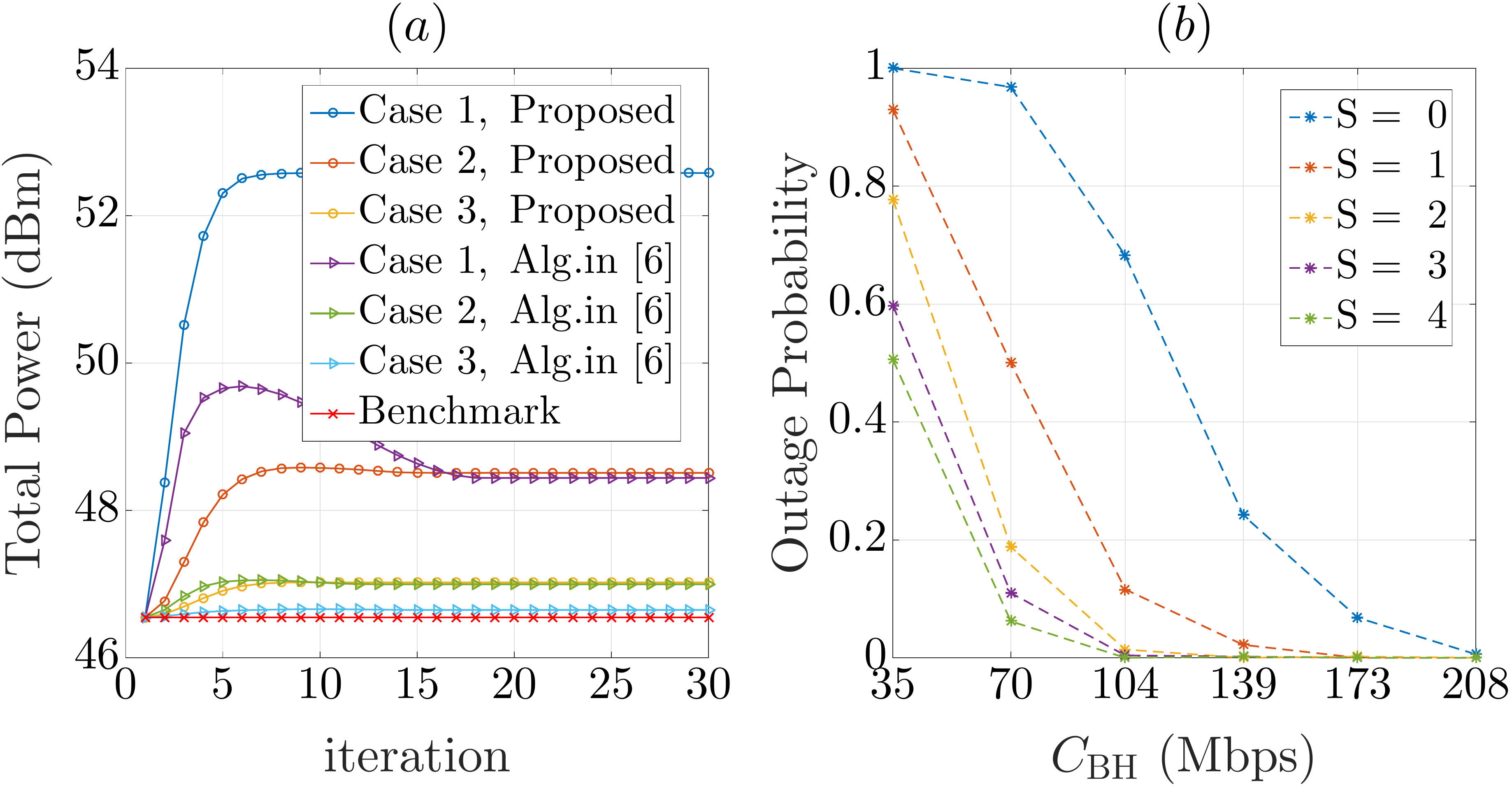}
	\caption{$(a)$ Comparison of total power consumption. \textit{Benchmark scheme: Full cooperation between all RUs for all multicast groups. Case 1: $70\ \text{Mbps},\ 2\ \text{Files Cached}$; Case 2: $104\ \text{Mbps},\ 2\ \text{Files Cached}$, Case 3: $104\ \text{Mbps},\ 3\ \text{Files Cached}$.} $(b)$ Outage probability for different scenarios.}\label{outage}
\end{figure}

In Fig. \ref{outage}$(a)$, we compare the total power consumption for different individual backhaul capacities \footnote{The algorithm in \cite{MeixiaCach} considers total capacity thus we consider the average individual capacity for a fair comparison.} and the number of requested files which have been cached. Result shows that the power consumption can be reduced either by caching more files or increasing the backhaul capacity, due to more cooperation becoming possible. Moreover, the power consumption of the proposed algorithm is always higher than that of \cite{MeixiaCach}, due to individual backhaul capacity constraints.  Then we set up 500 realizations, 12 different users with different requests are scheduled in each one. The proposed algorithm is used to solve each resultant problem. Some of them are infeasible with infinite power consumption as the solution. In these realizations, it is impossible to satisfy the QoS of all users with the current channel configurations and individual backhaul capacities. This is due either to many uncached files being requested, or to small individual backhaul capacities leading to little cooperation to counteract the bad channel condition. We compute the infeasible (outage) probability based on 500 realizations and then obtain Fig. \ref{outage}$(b)$. With more cache memory and larger backhaul capacity, outage probability can be reduced due to more cooperation being possible.
\section{Conclusion}
In this paper we propose an efficient algorithm to design the beamforming vectors and clustering pattern in the downlink of F-RAN. It balances the backhaul traffic according to individual backhaul capacities, guarantees the QoS of each user and minimizes the power consumption. Simulation results are in consistence with the theory and show that the optimal results can be obtained just after several iterations.

\renewcommand\refname{Reference}
\bibliographystyle{unsrt}
\bibliography{bib}

\end{document}